\documentclass[aps, twocolumn, superscriptaddress, longbibliography]{revtex4-1}
\usepackage[english]{babel}
\usepackage{graphicx}
\usepackage{stmaryrd}
\usepackage{amsmath, amsfonts, amssymb}
\usepackage{mathrsfs}
\usepackage{braket}
\usepackage{color}
\usepackage{xspace}
\usepackage{amscd}
\usepackage{latexsym}
\usepackage{diagbox}
\usepackage{bm}
\definecolor{lightblue}{rgb}{0.13, 0.26, 0.99}

\usepackage[
colorlinks=true,
urlcolor=blue,
citecolor=blue,
linkcolor=blue,
hyperfootnotes=false]{hyperref}

\allowdisplaybreaks

\begin{document}

\title{Solvable lattice model for (2+1)D bosonic topological insulator}
\author{Yusuke Horinouchi}
\affiliation{RIKEN Center for Emergent Matter Science (CEMS), Wako, Saitama 351-0198, Japan}
\email{yusuke.horinouchi@riken.jp}
\date{\today}
\begin{abstract}
We construct an exactly sovable commuting projector Hamiltonian for (2+1)D bosonic topological insulator which is one of symmetry-protected topological (SPT) phases protected by U(1) and time-reversal $\mathbb{Z}_2^T$ symmetry, where the symmetry group is U(1)$\rtimes\mathbb{Z}_2^T$. The model construction is based on the decorated domain-wall interpretation of the $E_{\infty}$-page of a spectral sequence of a cobordism group that classifies the SPT phases in question. 
We demonstrate nontriviality of the model by showing an emergence of a Kramers doublet when the system is put on a semi-infinite cylinder $(-\infty,0]\times S^1$ with an inserted $\pi$-flux.
The surface anomaly manifests itself as a non-onsite representation of the U(1)$\rtimes\mathbb{Z}_2^T$ symmetry. Anomaly matching on a boundary is discussed within a simple boundary theory.
\end{abstract}
\maketitle

\section{introduction}\label{sec:intro}
 The discovery of the electronic $\mathbb{Z}_2$ topological insulator \cite{PhysRevLett.95.226801, PhysRevLett.95.146802, PhysRevLett.96.106802, PhysRevB.75.121306, Konig766} provides an enormous impetus to the subject of symmetry-protected topological (SPT) phases, which have been one of central issues in condensed matter physics for a decade. The SPT phases are characterized by a gapped short-range entangled ground state which cannot be adiabatically deformed to a trivial product state or an atomic insulator in the presence of a certain symmetry \cite{PhysRevB.83.035107, PhysRevB.87.155114, Chen1604}. Consequently, if an SPT phase is put on a space with a boundary, degenerate boundary states emerge as long as the symmetry of the bulk is preserved. Many examples of SPT phases are discovered for both bosonic \cite{PhysRevB.81.064439, PhysRevB.83.035107, PhysRevB.87.155114, Chen1604, PhysRevLett.110.046801, PhysRevB.86.115109, PhysRevX.3.011016, kapustin2014anomalies, kapustin2014symmetry} and fermionic systems \cite{PhysRevB.78.195125, Ryu_2010, doi:10.1063/1.3149495, PhysRevB.83.075103, Freed2013, PhysRevB.90.115141, Wang629, Kapustin2015} through classification theories. 

 In this paper, we focus on the bosonic analog of the electronic $\mathbb{Z}_2$ topological insulator in (2+1)-dimensional spacetime. The bosonic topological insulator or the bosonic quantum spin-Hall insulator, which is introduced in Ref.~\cite{PhysRevB.87.155114}, is a (2+1)D bosonic SPT phase protected by U(1) and the time-reversal $\mathbb{Z}_2^T$ symmetry, where the symmetry group is U(1)$\rtimes\mathbb{Z}_2^T$ and the time-reversal operator squares to 1. For the the electronic $\mathbb{Z}_2$ topological insulator in class AII (the class with Pin${}^{\tilde{c}}_+$ structure, i.e., fermionic U(1)$\rtimes\mathbb{Z}_2^T$ symmetry), a simple lattice model is available and various physical systems have shown to support the phase  \cite{PhysRevLett.95.226801, PhysRevLett.95.146802, PhysRevLett.96.106802,  Bernevig1757, Konig766, PhysRevLett.100.236601}. On the other hand, a realization of the bosonic counterpart is still elusive even though bulk topological response and surface anomaly are extensively studied through the group-cohomology classification \cite{PhysRevB.87.155114, PhysRevB.89.035147}, the Chern-Simons $K$-matrix theory \cite{PhysRevB.86.125119}, non-linear sigma models \cite{PhysRevB.87.174412, PhysRevB.91.134404}, the bordism theory \cite{kapustin2014bosonic}, and a proposal for microscopic realization \cite{PhysRevLett.113.267206}. 
 
 In light of this situation, we construct an exactly solvable interacting lattice model for the bosonic topological insulator in the present papar as a starting point toward realization of the phase. We note that the authors of Ref.~\cite{PhysRevB.87.155114} discuss the construction of a lattice model for the bosonic topological insulator which is characterized by the Dijkgraaf-Witten response action classified by the cohomology group of $H^3({\rm U}(1)\rtimes \mathbb{Z}_2^T;U(1)_T)$. However, algebraic derivation of the 3-cocycle in $H^3({\rm U}(1)\rtimes \mathbb{Z}_2^T;U(1)_T)$, which is crucial for construction of a concrete model, is not explicitly performed in Ref.~\cite{PhysRevB.87.155114}, and furthermore, the model construction requires the group ring $\mathbb{C}[{\rm U(1)}\rtimes\mathbb{Z}_2^T]$ as a local Hilbert space for which we do not know corresponding physical degrees of freedom. Therefore, we here take a different strategy which is an extension of the Dijkgraaf-Witten theory: We employ decorated domain-wall construction of an SPT phase \cite{ddwchen}, in which we decorate a codimension-$q$ symmetry-breaking domain wall with ($d$-$q$)-dimensional SPT phases (see Sec.~\ref{sec:DDW} for details). As we demonstrate in the subsequent sections, the method of the decorated domain wall allows us to construct an exactly solvable lattice model whose physical degrees of freedom are spins (hardcore bosons). 
 To show that the constructed model indeed realizes the bosonic topological insulator phase, we examine a nontrivial topological response which is predicted in Ref.~\cite{PhysRevB.89.035147}. Namely, we demonstrate an emergence of a Kramers doublet on a boundary of the bosonic topological insulator, if the system is put on a semi-infinite cylinder which is penetrated by a $\pi$-flux of the U(1)-gauge field. We also explicitly derive a non-onsite representation of ${\rm U(1)}\rtimes\mathbb{Z}_2^T$ symmetry on the boundary, which is a manifestation of the surface anomaly. 
 
 We organize the paper in the following manner: In Sec.~\ref{sec:DDW}, we review the concept of the decorated domain-wall interpretation of cohomology groups that classify the SPT phases. In particular, we argue that the bosonic topological insulator can be interpreted as decorated domain walls using a spectral-sequence approach to the cobordism theory. Based on the decorated domain-wall interpretation of SPT phases, we then construct in Sec.~\ref{sec:latticemodel} a commuting projector Hamiltonian that realizes the bosonic topological insulator. The uniqueness of the ground state follows naturally from the structure of the constructed Hamiltonian. A idea behind the model construction is discussed by using a toy model. In Sec.~\ref{sec:topresp}, we show that the constructed lattice model indeed realizes the bosonic topological insulator by demonstrating a non-trivial topological response of the model. We put the constructed lattice model on a semi-infinite cylinder with an inserted $\pi$-flux, and show that the boundary of the cylinder supports a Kramers doublet.
 We also argue that the boundary of the semi-infinite cylinder exhibits a 't Hooft anomaly which manifests itself as a non-onsite representation of the U(1)$\rtimes\mathbb{Z}_2^T$ symmetry. In Sec.~\ref{sec:mathcing}, we discuss 't Hooft anomaly matching of the model. Within a simple surface theory, we show that the ground state is either gapless or breaks time-reversal symmetry, which indeed matches the anomaly of the U(1)$\rtimes\mathbb{Z}_2^T$ symmetry.


\section{Decorated domain wall interpretation}\label{sec:DDW}
Our model construction is based on a physical decorated domain-wall interpretation of the cohomology group that classifies the SPT phases in question. In this section, we first briefly review the concept of decorated domain-wall construction with a simple example of (1+1)D SPT with $\mathbb{Z}_2\times\mathbb{Z}_2$ symmetry, i.e. the Haldane phase. We then apply the scheme to the (2+1)D bosonic topological insulator protected by U(1)$\rtimes\mathbb{Z}_2^T$ symmetry.

\subsection{Warm up: (1+1)D SPT phase with $\mathbb{Z}_2\times\mathbb{Z}_2$ symmetry}\label{sec:haldane}
We first consider the (1+1)D SPT phase with $\mathbb{Z}_2\times\mathbb{Z}_2$ symmetry, which is classified by the cohomology group 
\begin{align}
	H^2(\mathbb{Z}_2^L\times\mathbb{Z}_2^R;{\rm U}(1))\simeq H^1\left(\mathbb{Z}_2^L;H^1(\mathbb{Z}_2^R;{\rm U}(1))\right)(\simeq\mathbb{Z}_2),\label{eq:haldaneDDW}
\end{align}
whose generator is the nontrivial SPT phase known as the Haldane phase protected by the dihedral symmetry. Here the labels $L$ and $R$ are introduced formally to distinguish the two $\mathbb{Z}_2$ symmetries. A physical interpretation of the cohomology group $H^1\left(\mathbb{Z}_2^L;H^1(\mathbb{Z}_2^R;{\rm U}(1))\right)$ on the right-hand side of Eq.~\eqref{eq:haldaneDDW} is given in Ref.~\cite{ddwchen}, which we review in the following, where the generator of the cohomology group $H^1\left(\mathbb{Z}_2^L;H^1(\mathbb{Z}_2^R;{\rm U}(1))\right)$ is directly related to the wave function of the Haldane phase.

Recall first that the coefficient $H^1(\mathbb{Z}_2^R;{\rm U}(1))$ in Eq.~\eqref{eq:haldaneDDW} classifies the (0+1)D SPT phases protected by $\mathbb{Z}_2^R$ symmetry. Namely, the two elements of $H^1(\mathbb{Z}_2^R;{\rm U}(1))\simeq \mathbb{Z}_2$ represent the trivial and non-trivial $\mathbb{Z}_2^R$-SPT phases which are the charges $0^{CR}$ and $1^{CR}$ of the $\mathbb{Z}_2^R$ symmetry, respectively. We thus obtain
\begin{align}
	H^1(\mathbb{Z}_2^R;{\rm U}(1))\simeq\mathbb{Z}_2\simeq \{\ket{0^{CR}},\ket{1^{CR}}\},
\end{align}
where the label $CR$ refers to the charge of the $\mathbb{Z}_2^R$ symmetry. Then, according to the standard algebraic definition of group cohomology, the generator of the cohomology group $H^1\left(\mathbb{Z}_2^L;H^1(\mathbb{Z}_2^R;{\rm U}(1))\right)$ is the following homogeneous cocycle:
\begin{align}
	&\omega: \mathbb{Z}_2^L \times \mathbb{Z}_2^L \rightarrow H^1(\mathbb{Z}_2^R;{\rm U}(1))\simeq \{\ket{0^{CR}},\ket{1^{CR}}\},\label{eq:cocycleHaldane}\\
	&\omega(g_0,g_1)=\left\{
		\begin{array}{l}
			\ket{1^{CR}}\mbox{  if ($g_0$,$g_1$)=(1,0) or (0,1),} \\
			\ket{0^{CR}}\mbox{  if ($g_0$,$g_1$)=(0,0) or (1,1).}
		\end{array}\right. \label{eq:cocycleHaldane2}
\end{align}

To clarify the relation between the cocycle $\omega$ and the SPT wave function of the Haldane phase, let us consider a spin-1/2 chain, where we identify the eigenvalues $1$ and $-1$ of the Pauli matrix $\sigma^z$ with the two elements $0$ and $1$ of the symmetry group $\mathbb{Z}_2^L$, respectively. Then the cocyle Eq.~\eqref{eq:cocycleHaldane} can be regarded as a function which maps nearby two spins to $\mathbb{Z}_2^R$ charges. Consequently, Eq.~\eqref{eq:cocycleHaldane2} signifies that the $\mathbb{Z}_2^R$-charge $\ket{1^{CR}}$ ($\ket{0^{CR}}$) is assigned to a bond of the spin chain if the two edges of the bond is occupied with spins in the opposite (the same) direction. In short, the nontrivial cocycle $\omega$ signifies that a nontrivial $\mathbb{Z}_2^R$-charge $\ket{1^{CR}}$ is decorated to each spin-domain wall. By taking an equal-weight superposition of the domain-wall configuration, we finally obtain the $\mathbb{Z}_2^L\times\mathbb{Z}_2^R$-symmetric SPT wave function. 

An exactly solvable lattice model which realizes the above decorated domain-wall wave function is the well-known cluster Hamiltonian:
\begin{align}
	&H=-\sum_{j} P_j-\sum_{j} Q_j,\\
	&P_j:=\frac{1}{2}(1+\sigma_{j}^z\tau_{j+1/2}^x\sigma_{j+1}^z),\\
	&Q_j:= \frac{1}{2}(1+\tau_{j-1/2}^z\sigma_{j}^x\tau_{j+1/2}^z),
\end{align}
where $\sigma$ and $\tau$ are Pauli matrices which are defined on vertices and links, respectively. The generators of $\mathbb{Z}_2^L\times \mathbb{Z}_2^R$ symmetry are given by the operators $\prod_j \sigma_j^x$ and $\prod_j \tau_{j+1/2}^x$. In the lattice model, all terms $\{P_j, Q_j\}_j$ are projection operators satisfying $P_j^2=P_j$ and $Q_j^2=Q_j$ and commute with one another. Therefore, every state in the entire Hilbert space is labeled uniquely by eigenvalues of the mutually-commuting projection operators $\{P_j, Q_j\}_j$ and, in particular, the state on which the operators $\{P_j, Q_j\}_j$ take the value of 1 is the unique gapped ground state $\ket{GS}$:
\begin{eqnarray}
	&\ket{GS}=\left(\prod_j Q_j\right)\ket{\cdots\rightarrow\uparrow\rightarrow\uparrow\cdots}\label{eq:gshaldane}\\
	&=\left(\prod_j \frac{1}{2}(1+\tau_{j-1/2}^z\sigma_{j}^x\tau_{j+1/2}^z)\right)\ket{\cdots\rightarrow\uparrow\rightarrow\uparrow\cdots},\label{eq:gshaldaneddw}
\end{eqnarray}
where $\ket{\cdots\rightarrow\uparrow\rightarrow\uparrow\cdots}$ represents the state with $\tau_{j+1/2}^x=1$ and $\sigma_j^z=1$ for all $j$. Due to the relation $Q_j^2=Q_j$, we can easily see that the right-hand side of Eq.~\eqref{eq:gshaldane} is the eigenvector of the operators $\{P_j, Q_j\}_j$ with the eigenvalues of 1. 

The right-hand side of Eq.~\eqref{eq:gshaldaneddw} directly leads to the decorated domain-wall interpretation. We first note that $\ket{\rightarrow}_{j+1/2}$ ($\ket{\leftarrow}_{j+1/2}$) is the trivial (nontrivial) (0+1)D SPT phase protected by $\mathbb{Z}_2^R$ symmetry which is generated by $\prod_j \tau_{j+1/2}^x$. Therefore, when acted on the state $\ket{\cdots\rightarrow\uparrow\rightarrow\uparrow\cdots}$,  the second term $\tau_{j-1/2}^z\sigma_{j}^x\tau_{j+1/2}^z$ of the operator $Q_j$ flips the spin $\sigma_j^z$ and decorates nearby bonds $j-1/2$ and $j+1/2$ with the nontrivial (0+1)D SPT phases $\ket{\leftarrow}_{j-1/2}$ and $\ket{\leftarrow}_{j+1/2}$. The action of $Q_j\propto 1+\tau_{j-1/2}^z\sigma_{j}^x\tau_{j+1/2}^z$ on the state thus creates a superposition of absence and presence of the spin domain at site $j$ with properly decorated $\mathbb{Z}_2^R$-SPT phases. Therefore, we can conclude that the product $\prod_i Q_i$ in Eq.~\eqref{eq:gshaldaneddw} produces an equal-weight superposition of all possible domain-wall configuration where the domain walls are decorated with the nontrivial (0+1)D $\mathbb{Z}_2^R$-SPT phases. 

We note that Ref.~\cite{PhysRevB.96.165124} clarifies the relation between the cluster Hamiltonian and the Haldane phase of the spin-1 chain.


\subsection{Bosonic topological insulator as decorated domain walls}\label{sec:ddwBTI}

Similarly to the (1+1)D $\mathbb{Z}_2\times\mathbb{Z}_2$ SPT phase, we can understand the (2+1)D bosonic topological insulator, which is an SPT phase protected by U(1)$\rtimes\mathbb{Z}_2^T$ symmetry, within the decorated domain-wall interpretation. According to the cobordism classification performed in Appendix~\ref{sec:classification}, the bosonic topological insulator is identified with the generator of the cohomology group of
\begin{align}
	H^2(\mathbb{Z}_2^T;H^2({\rm U}(1);\mathbb{Z}))\simeq \mathbb{Z}_2,
\end{align} 
which agrees with the previously obtained classification \cite{PhysRevB.87.155114,PhysRevB.89.035147,kapustin2014bosonic}.
On the left-hand side, the coefficient $H^2({\rm U}(1);\mathbb{Z})\simeq \mathbb{Z}$ is nothing but the (0+1)D SPT protected by the U(1) symmetry, i.e., the U(1) charge parametrized by an integer. 
According to a standard algebraic definition of group cohomology, the generator $\nu\in H^2(\mathbb{Z}_2^T;H^2({\rm U}(1);\mathbb{Z}))$ is represented by the following homogeneous cocycle:
\begin{align}
	&\nu:\mathbb{Z}_2^T\times \mathbb{Z}_2^T\times \mathbb{Z}_2^T\rightarrow H^2({\rm U}(1);\mathbb{Z})\simeq \mathbb{Z},\\
	&\nu(g_0,g_1,g_2)=\left\{
		\begin{array}{l}
			\ket{1^{C}}\mbox{  if ($g_0$,$g_1$,$g_2$)=(1,0,1) or (0,1,0),} \\
			\ket{0^{C}}\mbox{  otherwise,}
		\end{array}\right. \label{eq:nu}
\end{align}
where we introduce the label $C$ to denote the U(1) charge, i.e., $1^C$ refers to the U(1) charge of +1.

\begin{figure}[tb]
  	\begin{center}
  			\includegraphics[width=180pt]{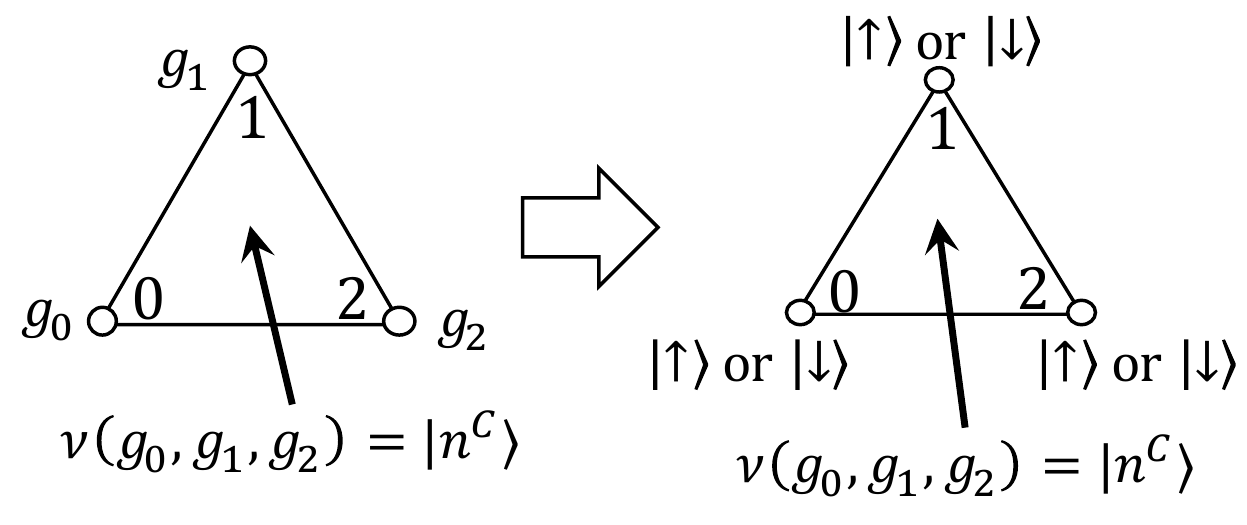}
			\caption{Graphical representation of the cocycle $\nu\in H^2(\mathbb{Z}_2^T;H^2({\rm U}(1);\mathbb{Z}))$, which maps three $\mathbb{Z}_2^T$ values $\{g_0,g_1,g_2\}$ to a U(1) charge parametrized by an integer $n^C$. As discussed in the main text, we identify the $\mathbb{Z}_2^T$ values on vertices with the physical degrees of freedom of spin-up and -down states $\{\ket{\uparrow},\ket{\downarrow}\}$.}\label{fig:ddwtriangle}
  	\end{center}
\end{figure}

Similarly to Sec.~\ref{sec:haldane}, the homogeneous cocycle $\nu$ can be interpreted as decorated domain walls (or more precisely decorated `defects'). 
To see this, let us consider a triangle on which $\mathbb{Z}_2^T$-values are assigned to the three vertices which are labeled by the numbers $i=$0, 1 and 2. Then the map $\nu$ is graphically represented as the left figure in Fig.~\ref{fig:ddwtriangle}, where the three $\mathbb{Z}_2^T$-values $(g_0,g_1,g_2)$ assigned to vertices map to $\ket{n^C}=\nu(g_0,g_1,g_2)$ at the center of the triangle. 
On each vertex $i\in\{0,1,2\}$, we then identify the two elements $0$ and $1$ of $\mathbb{Z}_2^T$ with the eigenvalues $1$ and $-1$ of the Pauli matrix $\sigma^z_i$ as depicted in the right figure of Fig.~\ref{fig:ddwtriangle}. 
Then the cocycle $\nu$ satisfying Eq.~\eqref{eq:nu} signifies that the center of the triangle should be decorated with a nontrivial U(1)-charge $\ket{1^{C}}$ if and only if the spin configuration is $(\sigma^z_0,\sigma^z_1,\sigma^z_2)=(1,-1,1)$ or $(-1,1,-1)$.
Similarly to Sec.~\ref{sec:haldane}, we can finally obtain the SPT wave function by taking a superposition of all possible spin configuration with appropriately decorated U(1) charges. 

\section{Lattice model for bosonic topological insulator}\label{sec:latticemodel}

To construct an SPT phase from the cocycle in Eq.~\eqref{eq:nu}, we consider the triangular lattice depicted in Fig.~\ref{fig:lattice}. 
On the lattice, a hardcore boson $\alpha$ with U(1) charge $+1$ is assigned to each up-pointing triangle and a hardcore boson $\beta$ with U(1) charge $-1$ is assigned to each down-pointing triangle.
For each vertex, we assign a pseudo spin-1/2 $\sigma$.
Here, $\sigma$ represents the Pauli matrices and $\alpha$ and $\beta$ are annihilation operators of hardcore bosons satisfying
\begin{align}
	&\{\sigma^x,\sigma^z\}=0, \\
	&(\sigma^x)^2=(\sigma^z)^2=1, \\
	&\{\alpha,\alpha^{\dag}\}=\{\beta,\beta^{\dag}\}=1,\\
	&\alpha^2=(\alpha^{\dag})^2=\beta^2=(\beta^{\dag})^2=0,\\
	&[\alpha_i,\alpha_j^{\dag}]=[\beta_i,\beta_j^{\dag}]=[\alpha_i,\alpha_j]=[\beta_i,\beta_j]=0 \mbox{ if }i\neq j,
\end{align}
where we note that a hardcore boson satisfies the fermionic anti-commutation relation on a single site.

\subsubsection{Symmetry}
As the U(1)$\rtimes\mathbb{Z}_2^T$ symmetry protecting the bosonic topological-insulator phase, we consider
\begin{align}
	&U(\theta)=\exp\left[i\theta\left(\sum_{\substack{{\rm up\mathchar`-pointing}\\{\rm triangles}}}\alpha^{\dag}\alpha-\sum_{\substack{{\rm down\mathchar`-pointing}\\{\rm triangles}}}\beta^{\dag}\beta\right)\right],\label{eq:Utheta}\\
	&T=\left(\prod_{\rm vertices}\sigma^x\right)\mathcal{K}\label{eq:Tr},
\end{align}
where $\theta\in\mathbb{R}/2\pi\mathbb{Z}$ and the anti-unitary operator $\mathcal{K}$ is the complex conjugation. We note that the minus sign in front of $\sum\beta^{\dag}\beta$ signifies that the hardcore boson $\beta$ has the U(1) charge of $-1$. It is straightforward to see that the time-reversal operator $T$ squares to 1 and that the symmetry group is U(1)$\rtimes\mathbb{Z}_2^T$, i.e., $T U(\theta)=U(-\theta) T$.

\subsubsection{SPT wave function}
According to the decorated domain-wall interpretation of the bosonic topological insulator discussed in Sec.~\ref{sec:ddwBTI}, the SPT wave function on the triangular lattice can be represented formally as
\begin{align}
	&\ket{{\rm SPT}}\nonumber\\
	&=\sum_{\substack{{\rm spin}\\{\rm configuration}}}\prod_{\substack{{\rm up\mathchar`-pointing}\\{\rm triangles}\:\Delta}}\nu(g_0,g_1,g_2)\nonumber\\
	&	\times\prod_{\substack{{\rm down\mathchar`-pointing}\\{\rm triangles}\:\nabla}}\left[-\nu(g_0',g_1',g_2')\right]\ket{{\rm spins}},\label{eq:wfddw}
\end{align}
which signifies that the SPT wave function is an equal-weight superposition of all possible spin configurations $\ket{\rm spins}$ with appropriately decorated U(1) charges $\nu$ in accord with Eq.~\eqref{eq:nu}.
We note that a down-pointing triangle is decorated with U(1) charge $-\nu$ with an additional minus sign. The minus sign assigned to the triangle with a reversed orientation is necessary so that the resultant wave function is U(1) symmetric. The same minus sign is needed for the Dijkgraaf-Witten theory developed in Ref.~\cite{PhysRevB.87.155114}.

\begin{figure}[tb]
  	\begin{center}
  			\includegraphics[width=240pt]{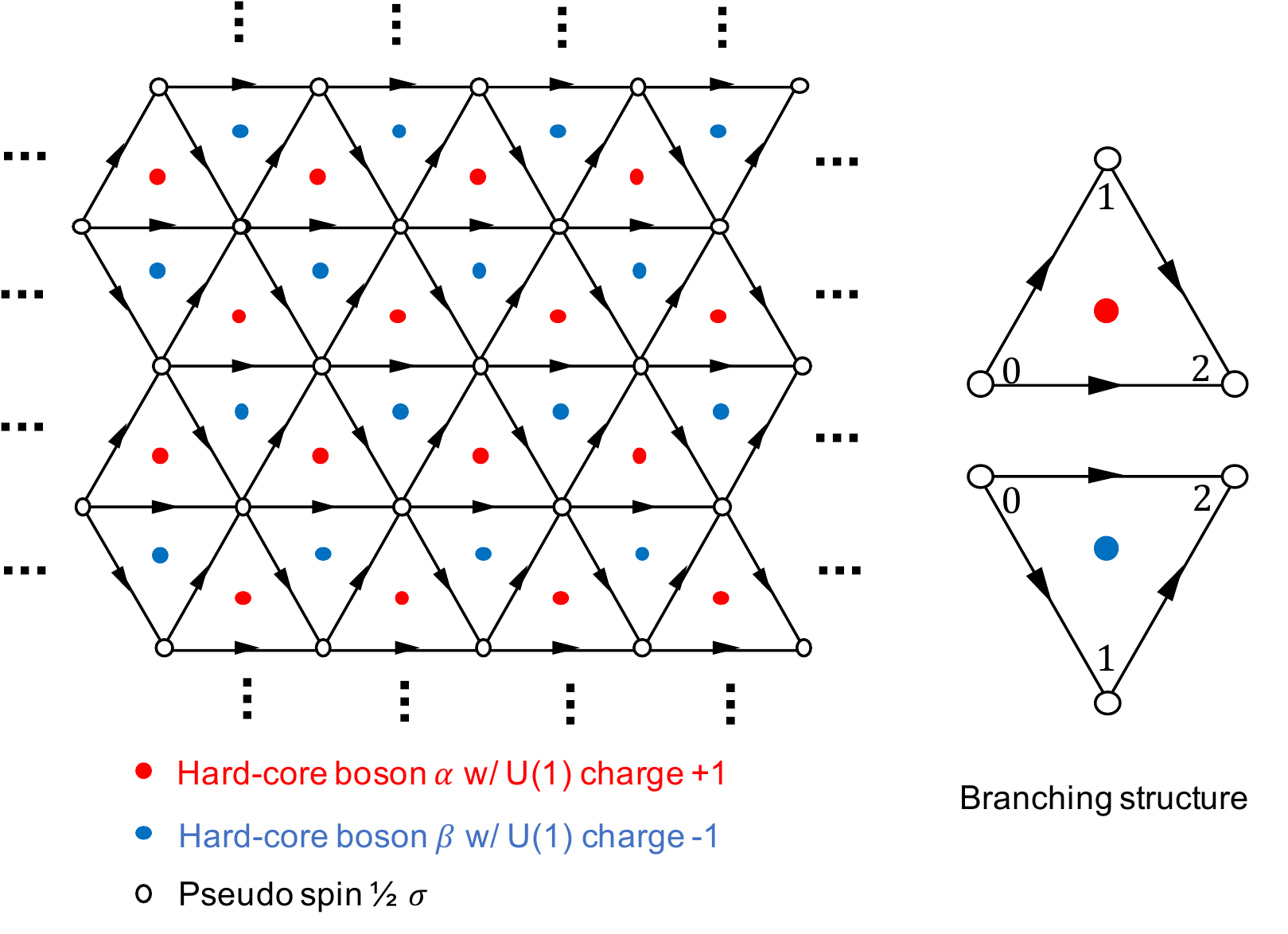}
			\caption{The lattice on which we construct the bosonic topological-insulator phase. A hardcore boson $\alpha$ with U(1) charge $+1$ is assigned to each up-pointing triangle. For a down-pointing triangle, we assign a hardcore boson $\beta$ with U(1) charge $-1$. For a vertex, we assign a pseudo spin-1/2 $\sigma$. As noted in the main text, U(1) charge $-1$ is necessary so that we construct a U(1)-symmetric SPT wave function. A spin $\sigma$ is called a `pseudo' spin since we consider the time-reversal operator $T$ which does not flip $\sigma^x$. The right figure specifies the labels 0, 1 and 2 assigned to the vertices of a triangle, i.e., the branching structure in other words. The branching structure is also represented by the arrows on the edges of a triangle.}\label{fig:lattice}
  	\end{center}
\end{figure}
 \begin{figure}[tb]
  		\begin{center}
  			\includegraphics[width=220pt]{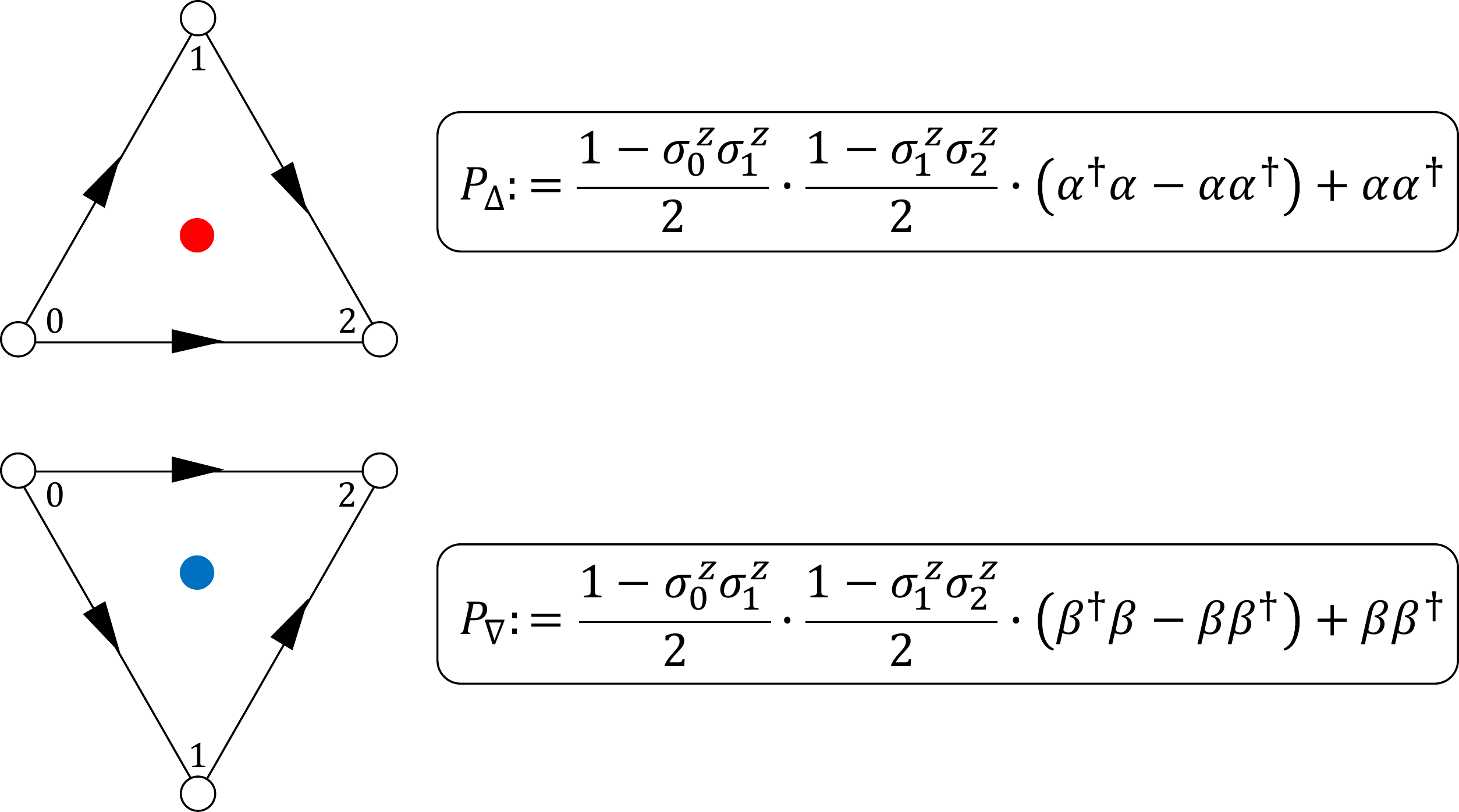}
  			\includegraphics[width=250pt]{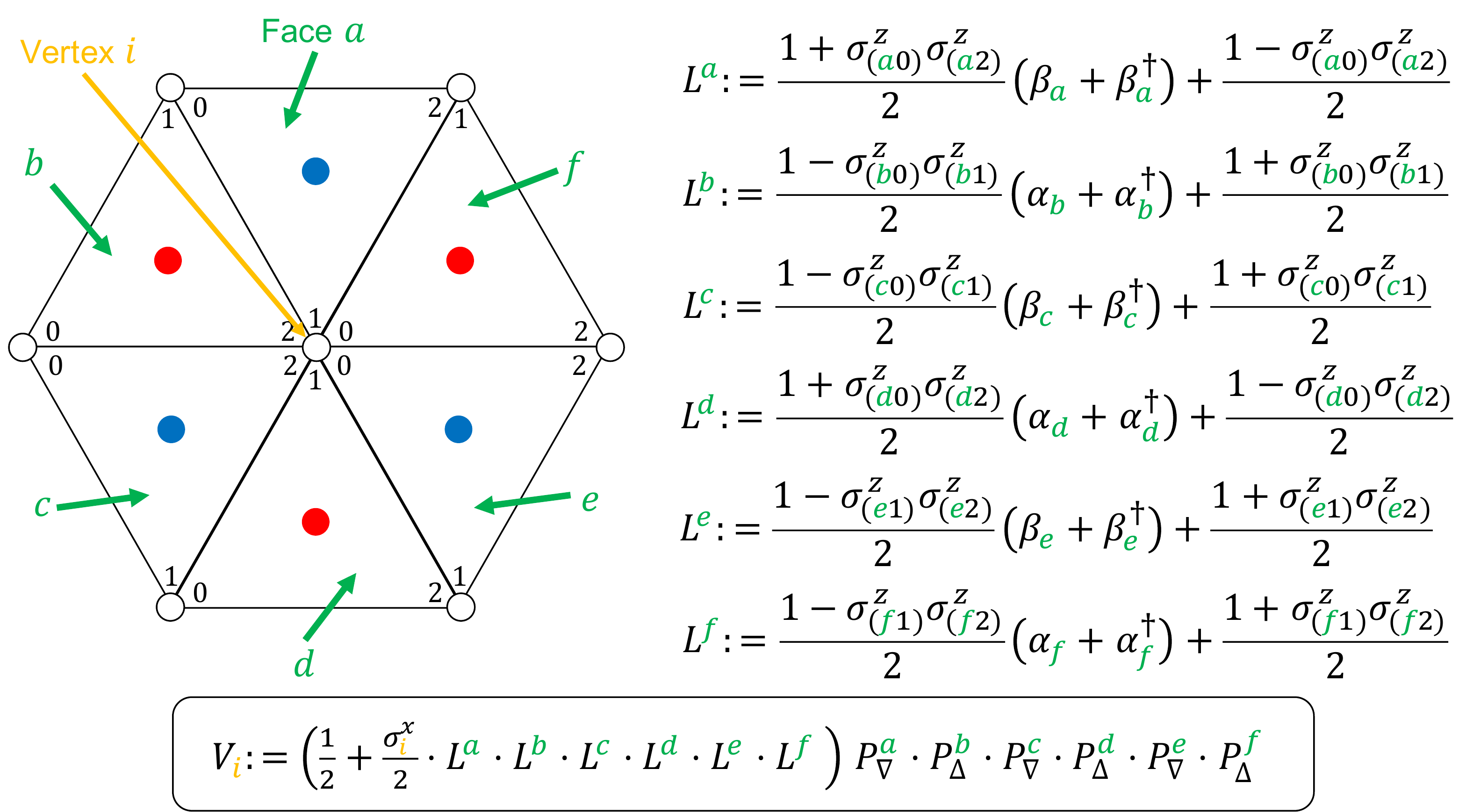}
			\caption{The operators appearing in the Hamiltonian Eq.~\eqref{eq:hamiltonian}. The operator $P_{\Delta}$($P_{\nabla}$) is defined for each up(down)-pointing triangle, and the operator $V_i$ is defined for each vertex. The operational meaning of these operators are presented in the main text. The operators are presented in Eqs.~\eqref{eq:Pup}-\eqref{eq:Lf}. }\label{fig:LCP}
  		\end{center}
\end{figure}

\subsubsection{Lattice model}
What we have to do next is to construct a lattice Hamiltonian whose ground state is given by Eq.~\eqref{eq:wfddw}. We first explicitly write down the Hamiltonian on the triangular lattice in Fig.~\ref{fig:lattice}:
\begin{align}
	H=-\sum_{\substack{{\rm up\mathchar`-pointing}\\{\rm triangles}\:\Delta}}P_{\Delta}-\sum_{\substack{{\rm down\mathchar`-pointing}\\{\rm triangles}\:\nabla}}P_{\nabla}
	-\sum_{{\rm vertices}\: i}V_i,\label{eq:hamiltonian}
\end{align}
where the operators $P_{\Delta}$, $P_{\nabla}$ and $V_i$ are defined in Fig.~\ref{fig:LCP}. The operators are written down as
\begin{align}
	&P_{\Delta}:=\frac{1-\sigma_0^z\sigma_1^z}{2}\frac{1-\sigma_1^z\sigma_2^z}{2}\left(\alpha^{\dag}\alpha-\alpha\alpha^{\dag}\right)+\alpha\alpha^{\dag},\label{eq:Pup}\\
	&P_{\nabla}:=\frac{1-\sigma_0^z\sigma_1^z}{2}\frac{1-\sigma_1^z\sigma_2^z}{2}\left(\beta^{\dag}\beta-\beta\beta^{\dag}\right)+\beta\beta^{\dag},\label{eq:Pdown}\\
	&V_i:=\left(\frac{1}{2}+\frac{\sigma_i^x}{2}L^aL^bL^cL^dL^eL^f\right) P_{\nabla}^aP_{\Delta}^bP_{\nabla}^cP_{\Delta}^dP_{\nabla}^eP_{\Delta}^f\label{eq:Vi},
\end{align}
where the subscripts $\{0,1,2\}$ in Eqs.~\eqref{eq:Pup} and \eqref{eq:Pdown} label the vertices of each triangle, or equivalently, refer to the branching structure. The subscripts $\{a,b,c,d,e,f\}$ in Eq.~\eqref{eq:Vi} label the six faces (triangles) sharing the vertex $i$. Operators $L$ are defined as 
\begin{align}
	&L^a=\frac{1+\sigma_{(a0)}^z\sigma_{(a2)}^z}{2}\left(\beta_a+\beta_a^{\dag}\right)+\frac{1-\sigma_{(a0)}^z\sigma_{(a2)}^z}{2},\label{eq:La}\\
	&L^b=\frac{1-\sigma_{(b0)}^z\sigma_{(b1)}^z}{2}\left(\alpha_b+\alpha_b^{\dag}\right)+\frac{1+\sigma_{(b0)}^z\sigma_{(b1)}^z}{2},\\
	&L^c=\frac{1-\sigma_{(c0)}^z\sigma_{(c1)}^z}{2}\left(\beta_c+\beta_c^{\dag}\right)+\frac{1+\sigma_{(c0)}^z\sigma_{(c1)}^z}{2},\\
	&L^d=\frac{1+\sigma_{(d0)}^z\sigma_{(d2)}^z}{2}\left(\alpha_d+\alpha_d^{\dag}\right)+\frac{1-\sigma_{(d0)}^z\sigma_{(d2)}^z}{2},\\
	&L^e=\frac{1-\sigma_{(e1)}^z\sigma_{(e2)}^z}{2}\left(\beta_e+\beta_e^{\dag}\right)+\frac{1+\sigma_{(e1)}^z\sigma_{(e2)}^z}{2},\\
	&L^f=\frac{1-\sigma_{(f1)}^z\sigma_{(f2)}^z}{2}\left(\alpha_f+\alpha_f^{\dag}\right)+\frac{1+\sigma_{(f1)}^z\sigma_{(f2)}^z}{2}\label{eq:Lf}.
\end{align}

Since the model is a bit complicated, let us first clarify the role of the operators $P_{\Delta}$, $P_{\nabla}$ and $V_i$ by considering a toy model which is simple but shares an essential feature with the Hamiltonian Eq.~\eqref{eq:hamiltonian}. As the toy model, we consider a four-dimensional Hilbert space whose orthonormal basis is given by $\{\ket{0},\ket{1},\ket{2},\ket{3}\}$. Here, the states $\{\ket{0},\ket{1}\}$ simulate spin configurations with appropriately decorated U(1) charges on the lattice in Fig.~\ref{fig:lattice} and the states $\{\ket{2},\ket{3}\}$ simulate the other states. Now, what we would like to do is to have a Hamiltonian whose unique ground state is given by an equal-weight superposition of $\ket{0}$ and $\ket{1}$, which simulates the SPT wave function on the lattice in Fig.~\ref{fig:lattice}. The states $\{\ket{2},\ket{3}\}$, which simulate the {\it inappropriately} decorated domain walls, should not be included in the ground state-wave function. For this purpose, we consider the following Hamiltonian $h$:
\begin{align}
	&h=-\left(\ket{0}+\ket{1}\right)\left(\bra{0}+\bra{1}\right),
\end{align}
which indeed has a unique ground state of $\ket{0}+\ket{1}$. An important observation is that the Hamiltonian $h$ can be divided in two parts:
\begin{align}
	&h=-\left(\ket{0}\bra{0}+\ket{1}\bra{1}\right)-\left(\ket{1}\bra{0}+\ket{0}\bra{1}\right),
\end{align}
where the first part $\ket{0}\bra{0}+\ket{1}\bra{1}$ is a projection operator to the sub-Hilbert space consisting only of $\ket{0}$ and $\ket{1}$, i.e., the appropriately decorated domain walls. The second part $\ket{1}\bra{0}+\ket{0}\bra{1}$ causes a transition between different decorated domain walls. In the original Hamiltonian in Eq.~\eqref{eq:hamiltonian}, the operators $P_{\Delta}$ and $P_{\nabla}$ play the role of the projection $\ket{0}\bra{0}+\ket{1}\bra{1}$, and the operator $V_i$ plays the role of the transition $\ket{1}\bra{0}+\ket{0}\bra{1}$. In this way, the Hamiltonian $H$ is constructed so that the ground state of $H$ is an equal-weight superposition of spin configurations with properly decorated U(1) charges.

Going back to the original Hamiltonian $H$ in Eq.~\eqref{eq:hamiltonian}, let us elaborate more on the physical meaning of the operators $\{P_{\Delta}, P_{\nabla},V_i\}_{\Delta, \nabla, i}$. The projection operators $P_{\Delta}$ and $P_{\nabla}$ in Eqs.~\eqref{eq:Pup} and \eqref{eq:Pdown} are rewritten as
\begin{align}
	&P_{\Delta}=\frac{1-\sigma_0^z\sigma_1^z}{2}\frac{1-\sigma_1^z\sigma_2^z}{2}\alpha^{\dag}\alpha\nonumber\\
	&+\left(1-\frac{1-\sigma_0^z\sigma_1^z}{2}\frac{1-\sigma_1^z\sigma_2^z}{2}\right)\alpha\alpha^{\dag},\\
	&P_{\Delta}=\frac{1-\sigma_0^z\sigma_1^z}{2}\frac{1-\sigma_1^z\sigma_2^z}{2}\beta^{\dag}\beta\nonumber\\
	&+\left(1-\frac{1-\sigma_0^z\sigma_1^z}{2}\frac{1-\sigma_1^z\sigma_2^z}{2}\right)\beta\beta^{\dag},
\end{align}
which clarify the operational meaning of $P_{\Delta}$ and $P_{\nabla}$. 
If $\sigma_0^z=-\sigma_1^z$ and $\sigma_1^z=-\sigma_2^z$, the operator $P_{\Delta}$ ($P_{\nabla}$) reduces to $\alpha^{\dag}\alpha$ ($\beta^{\dag}\beta$) that projects the hardcore boson $\alpha$ ($\beta$) to the state where the particle number equals 1. For the other spin configurations, $P_{\Delta}$ ($P_{\nabla}$) reduces to $\alpha\alpha^{\dag}$ ($\beta\beta^{\dag}$) that projects the hardcore boson $\alpha$ ($\beta$) to the state where the particle number equals 0. In short, $P_{\Delta}$ and $P_{\nabla}$ are the projection onto the states satisfying Eq.~\eqref{eq:nu}.
Concerning the operator $V_i$ in Eq.~\eqref{eq:Vi}, the term $\sigma_i^xL^aL^bL^cL^dL^eL^f$ flips the spin at the vertex $i$ and deforms the configuration of U(1) charges surrounding the vertex $i$ so that the resultant configuration of U(1) charges correctly decorate the spin configuration in accord with Eq.~\eqref{eq:nu}. For example, when $\sigma_{(a0)}^z=\sigma_{(a2)}^z$, the operator $L^a$ reduces to $\beta_a+\beta_a^{\dag}$ which flips the presence and the absence of the hardcore boson $\beta_a$ at the center of the triangle $a$, and when $\sigma_{(a0)}^z=-\sigma_{(a2)}^z$, the operator $L^a$ reduces to  $1$ which does not alter the configuration of $\beta_a$. The role of the term $P_{\nabla}^aP_{\Delta}^bP_{\nabla}^cP_{\Delta}^dP_{\nabla}^eP_{\Delta}^f$ in Eq.~\eqref{eq:Vi} is more subtle: The term guarantees that the operator $V_i$ is U(1) symmetric. Namely, the term $P_{\nabla}^aP_{\Delta}^bP_{\nabla}^cP_{\Delta}^dP_{\nabla}^eP_{\Delta}^f$ in the operator $V_i$ recovers the U(1) symmetry, which is apparently broken by the terms such as $\beta_a+\beta_a^{\dag}$ in $L^a$.

\subsubsection{Uniqueness of the ground state}
We finally discuss that the Hamiltonian $H$ in Eq.~\eqref{eq:hamiltonian} has a unique gapped ground state. The uniqueness of the ground state can be checked in the following manner:
\begin{enumerate}
\item Firstly, we can straightforwardly check that the operators $\{P_{\Delta}, P_{\nabla},V_i\}_{\Delta, \nabla, i}$ are mutually commuting projection operators satisfying $P_{\Delta}^2=P_{\Delta}$, $P_{\nabla}^2=P_{\nabla}$ and $V_i^2=V_i$. Namely, there is a basis that diagonalizes the operators $P_{\Delta}$, $P_{\nabla}$ and $V_i$ for all $\Delta$, $\nabla$ and $i$, and each eigenvalue of the operators is either 0 or 1.
\item Secondly, we can see that the number of the basis vectors equals to that of possible eigenvalues of the operators $\{P_{\Delta}, P_{\nabla},V_i\}_{\Delta, \nabla, i}$ due to the following three equations: \\(the number of bosons $\alpha$)=(the number of $P_{\Delta}$), \\(the number of bosons $\beta$)=(the number of $P_{\nabla}$), \\ (the number of spins $\sigma$)=(the number of $V_i$). 
\end{enumerate}
It then follows that each basis vector is {\it uniquely} labeled by the eigenvalues of the operators $\{P_{\Delta}, P_{\nabla},V_i\}_{\Delta, \nabla, i}$ and, in particular, the unique ground state is the state in which all the opeartors $\{P_{\Delta}, P_{\nabla},V_i\}_{\Delta, \nabla, i}$ take the eigenvalue 1. The excitation gap above the ground state is exactly 1 regardless of the system size since one of the operators in $\{P_{\Delta}, P_{\nabla},V_i\}_{\Delta, \nabla, i}$ takes on the value of 0 for the first excited state.

\section{Surface anomaly and topological response}\label{sec:topresp}
\begin{figure}[tb]
  		\begin{center}
  			\includegraphics[width=230pt]{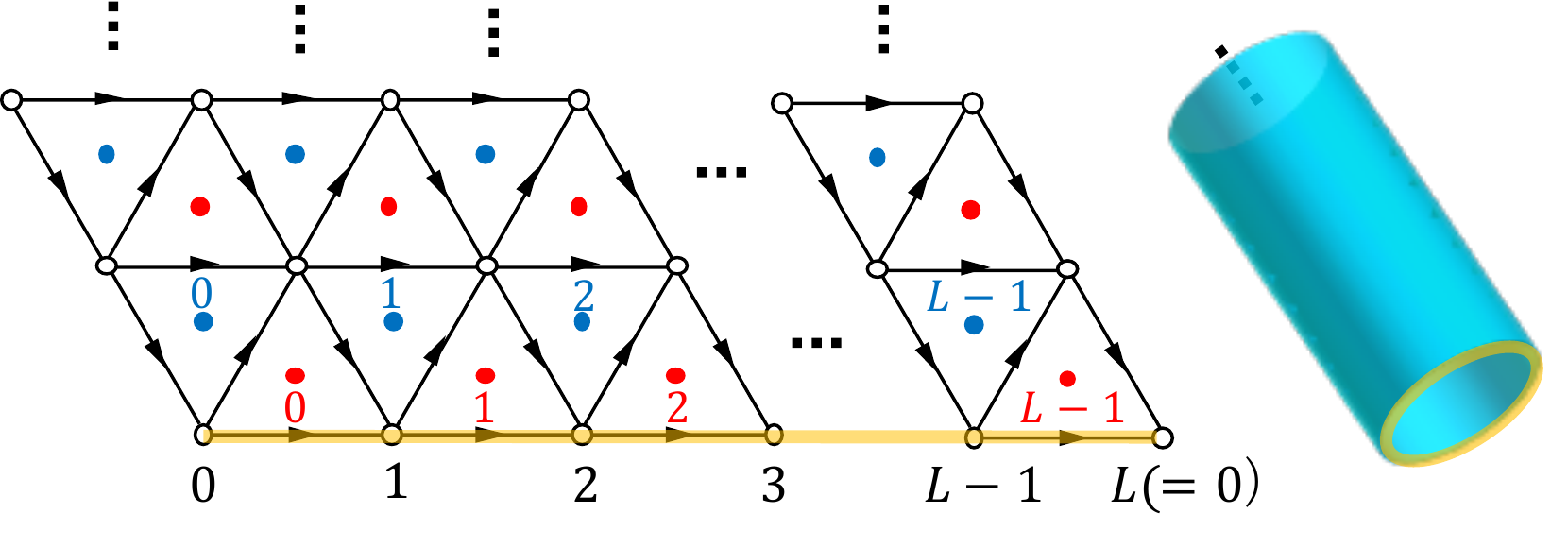}
			\caption{The semi-infinite cylinder on which we discuss the topological response and the surface anomaly. The yellow-colored curve is the boundary of the cylinder. Here $L$ refers to the system size of the boundary.}\label{fig:cylinder}
  		\end{center}
\end{figure}

In the last section, we construct a commuting-projector Hamiltonian from the decorated domain-wall interpretation of SPT phases. In this section, we argue that the constructed model indeed realizes the bosonic topological insulator. For this purpose, we consider the lattice model put on a semi-infinite cylinder as depicted in Fig.~\ref{fig:cylinder}. 
According to Ref.~\cite{PhysRevB.89.035147}, a Kramers doublet emerges on the boundary of the cylinder if a $\pi$-flux of the U(1) gauge field is inserted to the cylinder \cite{fn1}.
We thus examine the topological response within our lattice model.

\subsubsection{Surface degrees of freedom}
\begin{figure}[tb]
  		\begin{center}
  			\includegraphics[width=220pt]{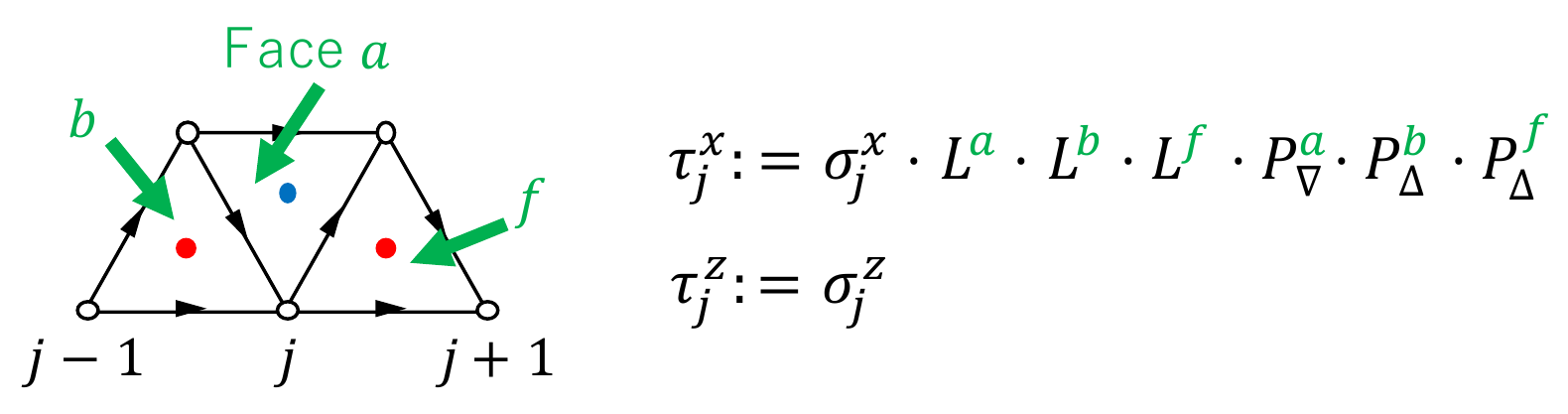}
			\caption{The dressed spin operator defined on the boundary of the semi-infinite cylinder. Here $j=0,1,\cdots,L-1$ labels the vertices on the boundary as in Fig.~\ref{fig:cylinder}. The operators $L$ and $P$ are defined in Eqs.~\eqref{eq:Pup}-\eqref{eq:Lf} or in Fig.~\ref{fig:LCP}.}\label{fig:dressedspin}
  		\end{center}
\end{figure}

For a later purpose, we first introduce `dressed' spin operators on the boundary of the semi-infinite cylinder depicted in Fig.~\ref{fig:cylinder}. Compared with the original boundary-less model, the Hamiltonian $H$ in Eq.~\eqref{eq:hamiltonian} put on the semi-infinite cylinder lacks the vertex terms $V_i$ on the boundary, since three triangles surrounding the vertex $i$ are absent.
Therefore, the Hamiltonian $H$ on the semi-infinite cylinder exhibits $2^L$-fold degenerate ground states which were originally labeled by the $2^L$ eigenvalues of the defunct vertex operators on the boundary. To label these $2^L$-fold degenerate ground states, we introduce dressed spin operators $\tau$ depicted in Fig.~\ref{fig:dressedspin}, which are written down as
\begin{align}
	&\tau_j^x:=\sigma_j^xL^aL^bL^fP_{\nabla}^aP_{\Delta}^bP_{\Delta}^f,\label{eq:taux}\\
	&\tau_j^z:=\sigma_j^z,\label{eq:tauz}
\end{align}
where the operators $L$ and $P$ are defined in Eqs.~\eqref{eq:Pup}-\eqref{eq:Lf} and the subscripts $\{a,b,f\}$ label the faces (triangles) sharing the vertex $j$. The operators $\tau_j^x$ and $\tau_j^z$ indeed represent a spin since they satisfy the algebra of the Pauli matrices:
\begin{align}
	&\left\{\tau_j^x,\tau_j^z\right\}=0,\\
	&\left(\tau_j^x\right)^2=\left(\tau_j^z\right)^2=1,
\end{align}
where we note that $\left(\tau_j^x\right)^2=P_{\nabla}^aP_{\Delta}^bP_{\Delta}^f=1$ holds since we now focus on the $2^L$-fold degenerate ground states on which $P_{\nabla}^a=P_{\Delta}^b=P_{\Delta}^f=1$. Other important properties of the dressed spin operators are that they commute with the bulk Hamiltonian and that they are mutually commuting for different vertices:
\begin{align}
	&\left[\tau_j^x,H\right]=\left[\tau_j^z,H\right]=0,\\
	&\left[\tau_j^x,\tau_l^x\right]=\left[\tau_j^z,\tau_l^z\right]=\left[\tau_j^x,\tau_l^z\right]=0\mbox{ if }j\neq l.
\end{align}
Therefore, the $2^L$-fold ground states are labeled uniquely by the eigenvalues of $\left\{\tau_j^z\right\}_j$ and $\tau_j^x$ flips the eigenvalue of $\tau_j^z$ without going outside the $2^L$-fold degenerate ground states.

\subsubsection{Anomalous representation of symmetry}
Based on the effective boundary degrees of freedom $\tau$, we deduce the representation of the U(1)$\rtimes\mathbb{Z}_2^T$ symmetry on the boundary. For this purpose, we first consider the action on $\tau$ of the symmetry generators $T$ and $U(\theta)$ which are defined in Eqs.~\eqref{eq:Utheta} and \eqref{eq:Tr}:
\begin{align}
	&T\tau_j^z T^{-1}=-\tau_j^z,\\
	&T\tau_j^x T^{-1}=\tau_j^x,\\
	&U(\theta)\tau_j^z U(-\theta)=\tau_j^z,\\
	&U(\theta)\tau_j^x U(-\theta)=\tau_j^x\exp\left[-i\frac{\theta}{2}\left(\tau_{j-1}^z\tau_j^z+\tau_j^z\tau_{j+1}^z\right) \right],
\end{align}
which can be obtained straightforwardly from Eqs.~\eqref{eq:taux}, \eqref{eq:tauz}, \eqref{eq:Utheta} and \eqref{eq:Tr}.
From these equations, we can deduce that the operators $T$ and $U(\theta)$ on the boundary can be represented effectively as
\begin{align}
	&T=\left(\prod_j \tau_j^x\right)\mathcal{K},\label{eq:TB}\\
	&U(\theta)=\exp\left(-i\frac{\theta}{2}\sum_j\frac{1-\tau_j^z\tau_{j+1}^z}{2}\right)\label{eq:UthetaB},
\end{align}
up to a phase factor. It is interesting to note that the representation Eq.~\eqref{eq:UthetaB} has a suggestive physical interpretation. Since the factor $\sum_j\frac{1-\tau_j^z\tau_{j+1}^z}{2}$ in $U(\theta)$ counts the number $N_{DW}$ of the spin-domain walls, $U(\theta)$ is represented as $e^{-i\frac{\theta}{2}N_{DW}}$, where due to the factor $-\frac{1}{2}$ in the exponent, each spin-domain wall has a fractionalized U(1) charge of $-\frac{1}{2}$.

Anomaly of the symmetry representation can be seen from the non-onsite nature of the operator $U(\theta)$ in Eq.~\eqref{eq:UthetaB}, i.e., $U(\theta)$ cannot be represented as a product of operators each of which is defined on a single site. The non-onsite nature prevents the U(1) symmetry on the boundary to be gauged \cite{PhysRevB.87.155114}, since a conventional gauging process requires an insertion of gauge fields on links between local degrees of freedom. Such an obstruction to gauging a symmetry is the manifestation of the 't Hooft anomaly \cite{PhysRevLett.112.231602}.

 \begin{figure}[tb]
  		\begin{center}
  			\includegraphics[width=220pt]{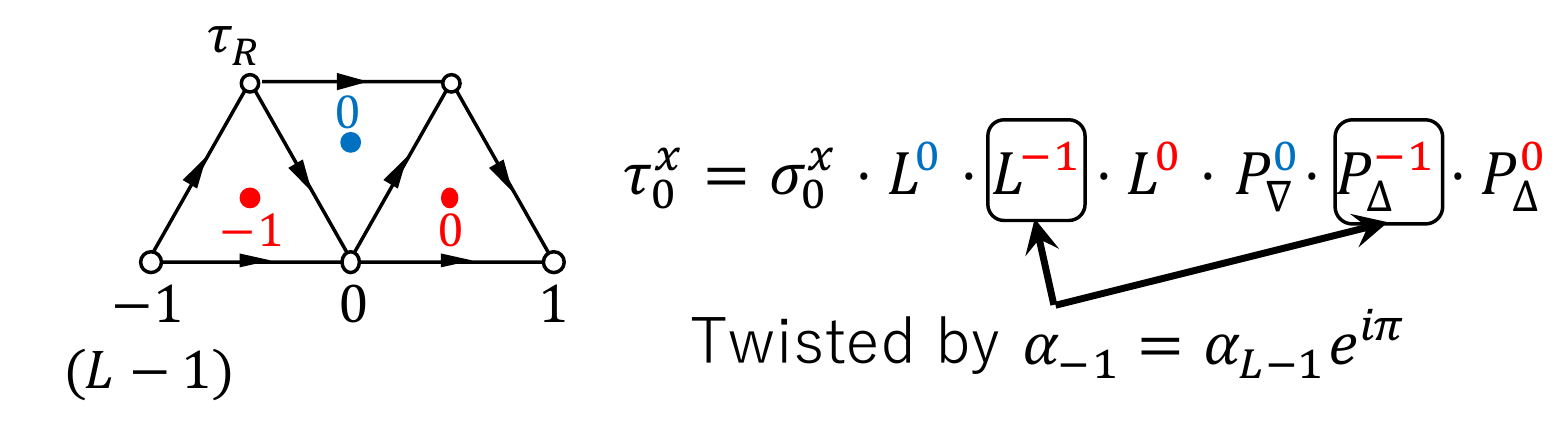}
			\caption{The effect of the twisted boundary condition Eqs.~\eqref{eq:tbcalpha} and \eqref{eq:tbcbeta}. While the periodic boundary condition imposes $\alpha_{-1}=\alpha_{L-1}$, the twisted boundary condition imposes $\alpha_{-1}=\alpha_{L-1}e^{i\pi}$. Consequently, $\alpha_{L-1}$ inside $\tau_x^0$ becomes $\alpha_{L-1}\rightarrow\alpha_{L-1}e^{i\pi}$ under the twisted boundary condition. Here, the spin degrees of freedom at the top-left vertex, which appears in Eq.~\eqref{eq:tbctau}, is represented by the Pauli matrix $\tau_R$.}\label{fig:twist}
  		\end{center}
\end{figure}

\subsubsection{Topological response}
We finally examine the effect of $\pi$-flux insertion to the semi-infinite cylinder. On the lattice in Fig.~\ref{fig:cylinder}, the $\pi$-flux insertion can be implemented as a twisted boundary condition along the tangential direction of the cylinder. Since the U(1) gauge field couple only to the hardcore bosons $\alpha$ and $\beta$ (Eq.~\eqref{eq:Utheta} consists of the hardcore bosons), the twisted boundary condition can be represented as 
\begin{align}
	&\alpha_{j+L}=U(\pi)\alpha_jU(-\pi)=\alpha_j\cdot e^{-i\pi}, \label{eq:tbcalpha}\\ 
	&\beta_{j+L}=U(\pi)\beta_jU(-\pi)=\beta_j\cdot e^{i\pi}, \label{eq:tbcbeta}
\end{align}
where the subscripts $j$ runs from 0 to $L-1$ as in Fig.~\ref{fig:cylinder}. On the boundary of the cylinder, the twisted boundary condition affects the effective surface degree of freedom $\tau^x_0$ introduced in Eq.~\eqref{eq:taux} which include $\alpha_{-1}$ in itself. As summarized in Fig.~\ref{fig:twist}, hardcore boson $\alpha_{L-1}$ included in $\tau^x_0$ is twisted by the $\pi$-flux, i.e., $\alpha_{L-1}\rightarrow \alpha_{L-1} e^{i\pi}$. Consequently, $\tau^x_0$ becomes
\begin{align}
	\tau^x_0\xrightarrow[\pi{\rm \mathchar`-flux}]{}\tilde{\tau}^x_0=\left(\tau_{L-1}^z\tau_R^z\right)\tau_0^x,\label{eq:tbctau}
\end{align}
where $\tau_R^z$ is the Pauli-$z$ matrix of a reference spin inside the bulk as depicted in Fig.~\ref{fig:twist}.

Now we are ready to show the emergence of the Kramers doublet under the twisted boundary condition. Due to Eq.~\eqref{eq:tbctau}, the time reversal operator $T$ introduced in Eq.~\eqref{eq:TB} becomes
\begin{align}
	&T=\left(\prod_{j=0}^{L-1}\tau_j^x\right)\mathcal{K}\nonumber\\
	&\xrightarrow[\pi{\rm \mathchar`-flux}]{}T_{\rm twist}=\tilde{\tau}^x_0\left(\prod_{j=1}^{L-1}\tau_j^x\right)\mathcal{K}=\left(\tau_{L-1}^z\tau_R^z\right)T.\label{eq:Ttwist}
\end{align}
While the original time-reversal operator $T$ squares to 1, the twisted time-reversal operator $T_{\rm twist}$ squares to $-1$, since on the right-hand side of Eq.~\eqref{eq:Ttwist}, $\tau_{L-1}^x$ included in $T$ and $\tau_{L-1}^z$ anti-commute with each other. The equation $T_{\rm twist}^2=-1$ signifies that the $\pi$-flux insertion to the semi-infinite cylinder produces a Kramers doublet on the boundary of the cylinder, which is the desired topological response of the bosonic topological insulator.

\section{Surface theory}\label{sec:mathcing}
Finally, we discuss the anomaly matching of a surface theory which preserves the U(1)$\rtimes\mathbb{Z}_2^T$ symmetry.
The 't Hooft anomaly matching of the U(1)$\rtimes\mathbb{Z}_2^T$ symmetry indicates that the ground states of a surface Hamiltonian $H_b$ are either gapless or breaking U(1)$\rtimes\mathbb{Z}_2^T$ symmetry, as long as $H_b$ commutes with the symmetry generators given in Eqs.~\eqref{eq:TB} and \eqref{eq:UthetaB}; however, we can not give a proof of the general statement so far. Therefore, we here discuss the anomaly matching in a more limited situation: As a surface theory, we consider a translationally invariant 3-local Hamiltonian $H_b$ which consists only of the terms containing nearby three spins. A possible Hamiltonian is given by
\begin{align}
	H_b=\sum_{j}\left(\tau_j^x-\tau_{j-1}^z\tau_j^x\tau_{j+1}^z\right)-h\sum_{j}\tau_j^z\tau_{j+1}^z.\label{eq:Hb}
\end{align}

\begin{figure}[tb]
  		\begin{center}
  			\includegraphics[width=230pt]{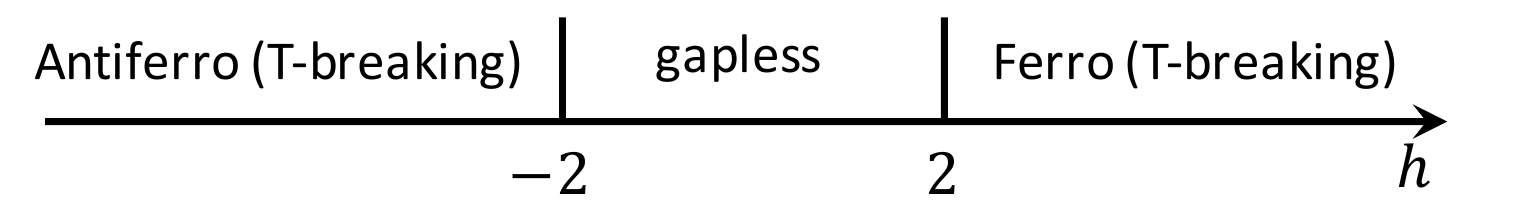}
			\caption{Phase diagram of the Hamiltonian $H_b$ in Eq.~\eqref{eq:Hb}. Time reversal-symmetry breaking phases, which support 2-fold degenerate ground states, emerge when $h>2$ and $h<-2$. A gapless phase emerges when $-2 \leq h \leq 2$.}\label{fig:pd}
  		\end{center}
\end{figure}

It is worth noting that the model is mapped to the XX-chain with a uniform magnetic field by the Kramers-Wannier transformation \cite{PhysRev.60.252} in the thermodynamic limit.
It is then natural to consider that the model can be solved analytically by means of the following Jordan-Wigner transformation:
\begin{align}
	&\gamma_j:=\left(\prod_{l=0}^{j-1}\tau_l^x\right) \tau_j^y,\\
	&\eta_j:=\left(\prod_{l=0}^{j-1}\tau_l^x\right) \tau_j^z,
\end{align}
where $\gamma_j$ and $\eta_j$ are Majorana fermion operators. Then the Hamiltonian $H_b$ in Eq.~\eqref{eq:Hb} becomes
\begin{align}
	H_b=-i\sum_{j=0}^{L-1}\left(\gamma_j\eta_j+\gamma_{j}\eta_{j+2}+h\gamma_j\eta_{j+1}\right),
\end{align}
where we impose $\gamma_{j+L}=-\gamma_{j}$ and $\eta_{j+L}=-\eta_{j}$ for the Hilbert space with even fermion parity and impose $\gamma_{j+L}=\gamma_{j}$ and $\eta_{j+L}=\eta_{j}$ for the Hilbert space with odd fermion parity.
Since $H_b$ is now the model of a free fermion, $H_b$ can be diagonalized analytically and we obtain the ground-state phase diagram as depicted in Fig.~\ref{fig:pd}. In the phase diagram, time reversal-symmetry-breaking ferromagnetic and anti-ferromagnetic phases emerge when $h>2$ and $h<-2$, respectively, where the ground-state energy is given by $-|h|L$. Here the ferromagnetic phase refers to the states in which the magnetization $\sum_j \tau_j^z$ takes on a nonzero value, and the anti-ferromagnetic phase refers to the states in which the Neel order parameter $\sum_j (-1)^j\tau_j^z$ takes on a nonzero value. The parameter region of $-2\leq h \leq 2$ supports a gapless phase where the dispersion relation of one-particle excitation is given by $\epsilon_k=2\left|2 \cos k+h\right|$, in which $k$ represents the momentum variable. Since the ground states are either gapless or break time-reversal symmetry, we can conclude that the Hamiltonian $H_b$ indeed matches the 't Hooft anomaly.

For a more generic Hamiltonian, we perform a numerical simulation; see Appendix~\ref{sec:dneq0}.

\section{Summary and discussion}
In the present paper, we construct an exactly solvable lattice model for the (2+1)D bosonic topological insulator which is an SPT phase protected by U(1)$\rtimes\mathbb{Z}_2^T$ symmetry. The discussion in the main text is summarized in the following. We first calculate the cobordism group that classifies the SPT phases protected by U(1)$\rtimes\mathbb{Z}_2^T$ symmetry by means of spectral sequence in algebraic topology. We then physically interpret the $E_{\infty}$-page of the spectral sequence as the decorated domain walls and construct the wave function of the bosonic topological insulator. An analysis of a toy model then enables us to construct an exactly solvable lattice model that realizes the decorated domain-wall wave function as the ground state. The topological non-triviality of the constructed model is then demonstrated by examining a topological response, which refers to an emergence of a Kramers doublet when the system is put on a semi-infinite cylinder penetrated by a $\pi$-flux. The surface anomaly of the U(1)$\rtimes\mathbb{Z}_2^T$ symmetry is shown by a non-onsite representation of the symmetry operators on a surface and the anomaly matching is discussed within a simple surface theory.

Our results may offer a numerical tool in searching for a system that realizes the bosonic topological insulator phase. Since we construct the SPT-wave function of the bosonic topological insulator on the lattice in Fig.~\ref{fig:lattice}, it is possible to measure numerically the overlap between the SPT-wave function and a numerically obtained ground-state wave function for another Hamiltonian on the lattice. By measuring the overlap, we may judge whether the bosonic topological insulator phase is realized or not as a ground state of a given Hamiltonian. 

A similar construction of exactly solvable lattice models may be possible for other SPT phases in which time-reversal defects are decorated with SPT phases in a lower dimension. For example, the (3+1)D bosonic topological insulator and the electric $\mathbb{Z}_2$ topological insulator in class AII are candidates of this construction, which may be addressed in a future publication.

\begin{acknowledgments}
The author acknowledges Yosuke Kubota, Shohei Miyakoshi, Takashi Mori and Ken Shiozaki for fruitful discussions. The author is grateful to Akira Furusaki and Shunsuke C. Furuya for their advices and comments on the manuscript. The author is supported by RIKEN Special Postdoctoral Researcher Program.
\end{acknowledgments}

\appendix

\section{Classification of U(1)$\rtimes\mathbb{Z}_2^T$ SPT phases}\label{sec:classification}
\begin{table}[b]
	\begin{tabular}{c|*{7}{c}}
		5 & 0 & 0 & 0 & 0 & 0 & 0  & 0    \\
		4 & 0 & $\mathbb{Z}_2$ & 0  & $\mathbb{Z}_2$ & 0  & $\mathbb{Z}_2$  & 0      \\
		3 & 0 & 0 & 0 & 0 & 0 & 0  & 0    \\
		2 & $\mathbb{Z}$ & 0  & $\mathbb{Z}_2$ & 0  & $\mathbb{Z}_2$     &  0    &   $\mathbb{Z}_2$    \\
		1 & 0 & 0 & 0 & 0 & 0 & 0  & 0  \\
		0 & 0& $\mathbb{Z}_2$ & 0  & $\mathbb{Z}_2$ & 0  & $\mathbb{Z}_2$     &  0       \\
		\hline
		\diagbox[dir=NE]{$q$}{$p$}
			&\makebox[3em]{0}&\makebox[3em]{1}&\makebox[3em]{2}&\makebox[3em]{3}
			&\makebox[3em]{4}&\makebox[3em]{5}&\makebox[3em]{6}\\
	\end{tabular}
	\caption{The $E_2$ page of the Leray-Serre spectral sequence Eq.~(\ref{eq:sss}). $p+q$ corresponds to $d$.}\label{tbl:E2pagesss}
\end{table}  

Here we classify the SPT phases protected by U(1)$\rtimes\mathbb{Z}_2^T$ symmetry. For the classification, we employ the Anderson dual $D\Omega_{{\rm O}\ltimes{\rm U}(1)}^d$ of the bordism groups with O($d$)$\ltimes$U(1) structure \cite{FH16}. A spectral sequence that converges to the cohomology group is \cite{may2004parametrized}
\begin{align}
	D\Omega_{{\rm O}\ltimes{\rm U}(1)}^d\Leftarrow E_2^{p,q}=H^p(S^{\infty}\times_{\mathbb{Z}_2}\mathbb{C}P^{\infty};\underline{D\Omega_{SO}^q(pt)}_T),
\end{align}
where $\underline{D\Omega_{SO}^q(pt)}_T$ is the Anderson dual  of the oriented bordism group on which $\mathbb{Z}_2^T$ acts nontrivially. Here $S^{\infty}$ denotes the infinite sphere and $\mathbb{C}P^{\infty}$ is the infinite union of the complex projective spaces. The space $S^{\infty}\times_{\mathbb{Z}_2}\mathbb{C}P^{\infty}$ is the classifying space of U(1)$\rtimes\mathbb{Z}_2^T$ where on $S^{\infty}$ and $\mathbb{C}P^{\infty}$, $\mathbb{Z}_2^T$ acts as the antipodal map and complex conjugation, respectively. As long as we are interested in the (2+1)D system, we have only to consider the case of $q=0$, since the oriented cobordism groups $D\Omega_{SO}^d=0$ for $d=1,2,3$. We thus employ the following Leray-Serre spectral sequence:
\begin{eqnarray}
	&&H^d\left(S^{\infty}\times_{\mathbb{Z}_2}\mathbb{C}P^{\infty};\underline{D\Omega_{SO}^0(pt)}_T\right)\nonumber\\
	&&\Leftarrow E_2^{p,q}=H^p\left(\mathbb{R}P^{\infty};\underline{H^q(\mathbb{C}P^{\infty};\mathbb{Z})}_{T_q}\otimes_{\mathbb{Z}} \underline{D\Omega_{SO}^0(pt)}_T\right),\nonumber\\
	\label{eq:sss}
\end{eqnarray}
where $T_q$ means that $\mathbb{Z}_2^T$ acts on $H^q(\mathbb{C}P^{\infty};\mathbb{Z})$ nontrivially (trivially) when $q=2$ ($q=0,4$): The complex conjugation changes (preserves) the orientation when $q=2$ ($q=0,4$). We summarize the $E_2$ page in Table~\ref{tbl:E2pagesss}. By comparing the obtained $E_2$ page and the results in Ref.~\cite{cadek1999}, we can see that the $E_2$ page survives up to the $E_{\infty}$-page. We thus obtain
\begin{align}
	D\Omega_{{\rm O}\ltimes{\rm U}(1)}^4\simeq H^2(\mathbb{R}P^{\infty};H^2(\mathbb{C}P^{\infty};\mathbb{Z})),
\end{align}
which agrees with Refs.~\cite{cadek1999,kapustin2014bosonic}. Due to the relations $\mathbb{R}P^{\infty}=B\mathbb{Z}_2$ and $\mathbb{C}P^{\infty}=B$U(1), we finally obtain
\begin{align}
	D\Omega_{{\rm O}\ltimes{\rm U}(1)}^4\simeq H^2(\mathbb{Z}_2;H^2(U(1);\mathbb{Z})).
\end{align}
As a consequence, the bosonic (2+1)D topological insulator can be constructed by decorating a codimension-2 defect with a U(1) charge. We note that the bosonic (3+1)D topological insulator can be constructed by decorating a codimension-1 domain wall with a bosonic integer quantum Hall system.

\section{Anomaly matching in a more generic Hamiltonian}\label{sec:dneq0}
 \begin{figure*}[t!]
  		\begin{center}
  			\includegraphics[width=500pt]{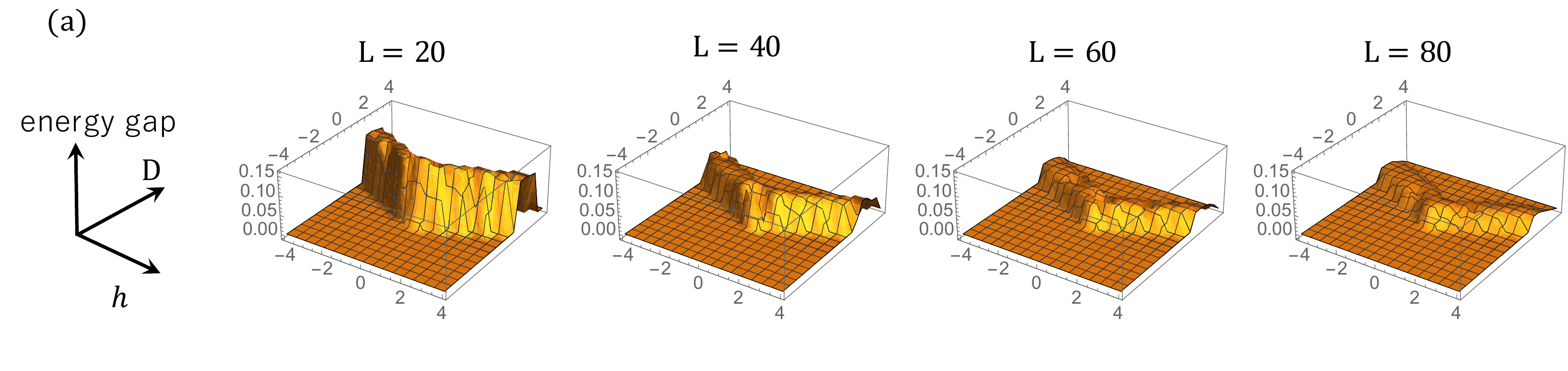}
			\includegraphics[width=500pt]{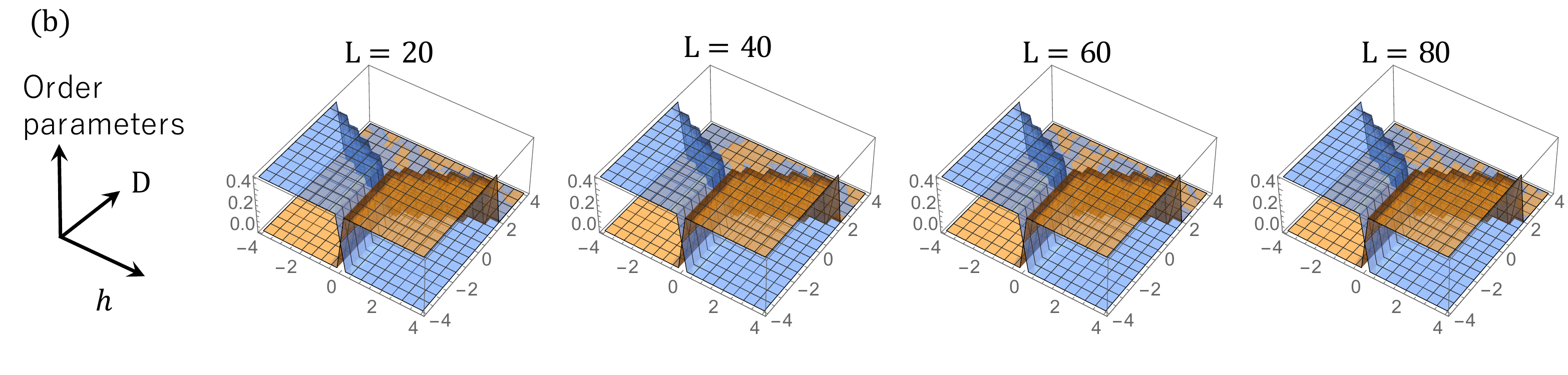}
			\caption{(a) Energy gap is plotted against the parameter $h$ and $D$ of the surface Hamiltonian Eq.~\eqref{eq:Hbg}. Here $L$ represents the system size of the (1+1)D boundary of the semi-infinite cylinder. In the region of $D\lesssim 0$, we see a large plateau where the energy gap is almost 0. In the region of $D\gtrsim 0$, we see a finite energy gap above a ground state which tends to decrease as the system size $L$ grows up. (b) The Neel order parameter (blue-colored surface) and the ferromagnetic order parameter (orange-colored surface) are plotted against the parameter $h$ and $D$ of the surface Hamiltonian. In the region of $D\lesssim 0$, where the energy gap is almost 0, the ground states exhibit the Neel order when $h<0$ and exhibit the ferromagnetic order when $h>0$.}\label{fig:energygap}
  		\end{center}
\end{figure*}
 \begin{figure*}[t!]
  		\begin{center}
  			\includegraphics[width=500pt]{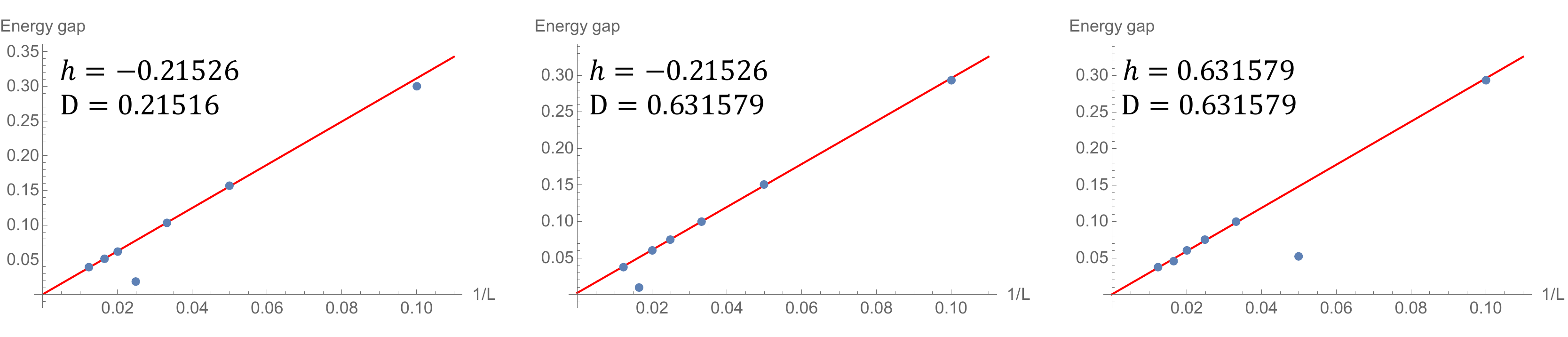}
			\caption{Energy gap is plotted against the inverse system size $1/L$ for some fixed parameters $h$ and $D$ of the surface Hamiltonian Eq.~\eqref{eq:Hbg}. We can see that the energy gap is proportional to $1/L$ except for some points. The red line shows the extrapolation without the exceptional points and the extrapolation leads to the energy gap less than $10^{-2}$ in the thermodynamic limit.}\label{fig:extrapolation}
  		\end{center}
\end{figure*}

This appendix complements the discussion in Sec.~\ref{sec:mathcing}. In particular, we discuss the surface-anomaly matching of a more generic translationally invariant 3-local Hamiltonian given by
\begin{align}
	H_b=\sum_{j}\left(\tau_j^x-\tau_{j-1}^z\tau_j^x\tau_{j+1}^z\right)-h\sum_{j}\tau_j^z\tau_{j+1}^z+D\sum_{j}\tau_j^z\tau_{j+2}^z.\label{eq:Hbg}
\end{align}
For the Hamiltonian, we numerically calculate the energy gap above a ground state, which is expected to be 0 in the thermodynamic limit due to the anomaly. For the purpose, we employ the density matrix-renormalization group (DMRG) method which are performed using the ITensor library \cite{Itensor}. We should note that the obtained results do not convincingly show that the energy gap becomes 0 in the thermodynamic limit. However, we explicitly present the results in this appendix since the results are suggestive enough.

Before we present the results, we here summarize the details of the parameters used in the DMRG calculation. For a fixed system size $L$ and the parameters $h$ and $D$ of the Hamiltonian, we calculate the energies of the ground and the first excited states, where the number of sweeps is 50, the maximum truncation error is $10^{-10}$ and the bond dimension is at most 1200. To perform the DMRG calculation under the periodic boundary condition, we employ the prescription of mapping a periodic spin chain to an open spin ladder: Site $i$ is re-labeled as $2i+1$ when $i\leq L/2$ and is re-labeled as $2(L-i-1)$ when $i>L/2$.

With these setup, we obtain Fig.~\ref{fig:energygap}(a), where the energy gap is plotted against the parameters $h$ and $D$ of the Hamiltonian in Eq.~\eqref{eq:Hbg} for an increasing system size $L$. 
We can see a large plateau in the region of $D\lesssim 0$, where 2-fold degenerate ground states emerge. In this region, the ground states spontaneously break the time-reversal symmetry and exhibit the Neel order when $h<0$ and exhibit the ferromagnetic order when $h>0$ as depicted in Fig.~\ref{fig:energygap}(b). In the region of $D\gtrsim 0$, we see a finite energy gap above a ground state which tends to decrease as the system size $L$ grows up; however, we could not perform a well-controlled extrapolation of the energy gap to the thermodynamic limit in this parameter region. To see this, we explicitly show in Fig.~\ref{fig:extrapolation} the extrapolation of the energy gap. As we can see, except for some points, the energy gap is proportional to the inverse system size $1/L$, which is expected for the gapless phase. The red line shows the extrapolation without the exceptional points and the extrapolation leads to the energy gap less than $10^{-2}$ in the thermodynamic limit. The result seems to suggest that the system supports the gapless phase but we can not reach the conclusion since we are not aware of the origin of the exceptional points.

In summary, we obtain the result which suggests that the energy gap in the thermodynamic limit is likely to be 0, but there appear some exceptional points obstructing the extrapolation of the energy gap to the thermodynamic limit. The origin of the exceptional points is not clear.

\bibliography{ref.bib}

\begin{thebibliography}{39}
\bibitem [{fn1()}]{fn1}%
  \BibitemOpen
  \href@noop {} {\enquote {\bibinfo {title} {The topological response is
  deduced from the fact that the bosonic topological insulator is nontrivial
  also as an {SPT} phase protected by $\mathbb{Z}_2\times\mathbb{Z}_2^{T}$
  symmetry, where $\mathbb{Z}_2=\{-1,1\}$ is a subgroup of {U}(1). {T}he
  topological action of the {SPT} phase is given by $\int_{M} w_1^2\smile a$,
  where $w_1$ is the first {S}tiefel-{W}hitney class defined on the tangent
  bundle of the spacetime ${M}$, and $a$ represents the $\mathbb{Z}_2$ gauge
  field. {A}s an {SPT} phase protected by $\mathbb{Z}_2\times\mathbb{Z}_2^{T}$
  symmetry, the $\pi$-flux is a nontrivial holonomy of the $\mathbb{Z}_2$ gauge
  field $a$ along the tangential direction of the cylinder, and is represented
  by the boundary condition twisted by the $\mathbb{Z}_2$ symmetry. {O}n the
  (1+1)-dimensional twisted boundary, the topological action then becomes that
  of the {H}aldane phase protected by $\mathbb{Z}_2^{T}$ symmetry. {A}s is well
  known, the {H}aldane phase supports a {K}ramers doublet on the boundary.}}\
  }\BibitemShut {NoStop}%
\end{thebibliography}

\end{document}